\documentclass{birkjour}
\usepackage{bm}
\usepackage{amsmath}
\usepackage{amsfonts}
\usepackage{amssymb}
\usepackage{enumerate}

\begin{document}

\title{Combinatorial Approach to Boson Anti-Normal Ordering Problem}
\author{M. R. Bazrafkan}
\address{Physics Department, Faculty of Science, I. K. I. University, Qazvin, Iran.}
\email{bazrafkan@ikiu.ac.ir}
\author{F. Sh\"ahandeh$^*$}
\address{Physics Department, Faculty of Science, I. K. I. University, Qazvin, Iran.}
\email{shahandeh@ikiu.ac.ir}
\author{E. Nahvifard}
\address{Physics Department, Faculty of Science, I. K. I. University, Qazvin, Iran.}
\email{nahvifard@ikiu.ac.ir}

\begin{abstract}
We address a systematic combinatorial approach to the anti-normal ordering problem. In this way, we use the Stirling numbers and their generating function, the so-called Bell polynomials, together with the operational methods to anti-normal the operator ${e^{\lambda {a^\dag }a}}$. In fact, we exploit the new theorem given by Sh\"ahandeh \emph{et~al.} in a special case of anti-normal ordering.
\end{abstract}

\thanks{$^*$Corresponding author.}


\maketitle

\section{Introduction} \label{Intro}

The combinatorial aspects of normal ordering procedure was first introduced by Katriel~\cite{Katriel} and then used systematically by Blasiak \emph{et~al.}~\cite{Blasiak1}. In general, they have introduced a combinatorial realization of the Heisenberg--Weyl algebra defined as a linear vector space $\mathcal{A}$ over some field $\mathbb{K}$ (see~\cite{Blasiak2}.) However, we ignore the details and restrict ourselves to a special example to show the main idea.~\cite{Blasiak3}

\sloppy
As a widely applicable example, consider we are going to normally order the monomial ${\left( {{a^\dag }a} \right)^n}$. Writing this as a $n$-term multiplicative sequence $\left( {{a^\dag }a} \right)\left( {{a^\dag }a} \right)\left( {{a^\dag }a} \right) \cdots \left( {{a^\dag }a} \right)$, Blasiak \emph{et~al.} in Ref.~\cite{Blasiak3} have shown that one may consider contractions as connectors between different blocks;
	\begin{equation*}
	\left( {{a^\dag }a} \right)\underbrace {\left( {{a^\dag }a} \right)\left( {{a^\dag }a} \right)} \cdots \left( {{a^\dag }a} \right) = \left( {{a^\dag }a} \right) \left( {{a^\dag }} \right)\left( a \right) \cdots \left( {{a^\dag }a} \right)~.
	\end{equation*}
In this way, connecting each two blocks by removing an operator $a$ standing on the left of an operator $a^\dag$ might be realized as putting two objects in one container. Thus, the number of $k$-pair contractions between these blocks is given by the number of ways putting $n$ different objects in $n-k$ identical containers. This is of course the number of $n-k$ partitions of a set with $n$ elements given by the familiar \emph{Stirling numbers of the second kind} $S\left(n,n-k\right)$.

Now, a simple calculation using the Wick's theorem~\cite{Navon} shows that one may expand the monomial ${\left( {{a^\dag }a} \right)^n}$ in a normally ordered power series as~\cite{Katriel,Blasiak3}
	\begin{equation*}
	{\left( {{a^\dag }a} \right)^n} = \sum\limits_{k = 1}^n {S\left( {n,k} \right):{{\left( {{a^\dag }a} \right)}^k}:}  = \sum\limits_{k = 1}^n {S\left( {n,k} \right){a^{\dag k}}{a^k}}~,
	\end{equation*}
in which $:\quad :$ is the (Wick's) normal ordering symbol. Exploiting generating function of Stirling numbers, the so called \emph{Bell polynomials}
	\begin{equation} \label{Bell}
	B\left( {n,x} \right) = \sum\limits_{k = 1}^n {S\left( {n,k} \right){x^k}}~,
	\end{equation}
one has
	\begin{equation*}
	{\left( {{a^\dag }a} \right)^n} = :B\left( {n,{a^\dag }a} \right):~.
	\end{equation*} 
Indeed, she may go further and use the generating function of Bell polynomials,
	\begin{equation} \label{Dobinski}
	G\left( {x,\lambda } \right) = \sum\limits_{n = 0}^\infty  {\frac{{{\lambda ^n}}}{{n!}}B\left( {n,x} \right)}  = e^{\left( {{e^\lambda } - 1} \right)x}~,
	\end{equation}
to write
	\begin{eqnarray} \label{sm}
	{e^{\lambda {a^\dag }a}} &=& \sum\limits_{n = 0}^\infty  {\frac{{{\lambda ^n}}}{{n!}}{{\left( {{a^\dag }a} \right)}^n}} 		\nonumber\\
	&=& :\sum\limits_{n = 0}^\infty  {\frac{{{\lambda ^n}}}{{n!}}B\left( {n,{a^\dag }a} \right)} : \nonumber\\
	&=& :G\left( {\lambda ,{a^\dag }a} \right): \nonumber\\
	&=& :e^{\left( {{e^\lambda } - 1} \right){a^\dag }a}:~.
	\end{eqnarray}

As we see, representing quantum operators by means of combinatorial objects leads to a simple and sensible demonstration of the process of normal ordering. Moreover, such a realization allows one to apply theorems and methods regarding discrete mathematics such as calculus of generating functions to the problem at hand.

\section{Combinatorics of the General Class of Orderings} \label{CGCO}

The general class of orderings, or $s$-ordering have been introduced by Cahill and Glauber in late 60's.~\cite{CG1,CG2} There exist two general procedures of $s$-ordering some operator function which both exploit integral transformations~\cite{Glauber,Fan}, however, Sh\"ahandeh and Bazrafkan have recently given a new third method of general ordering~\cite{SB} reducing the $s$-ordering problem to a purely combinatorial one. In this regard, they have shown that any multiplicative sequence of $s_j$-ordered functions, $j \in \left\{ {2,3, \ldots ,k} \right\}$ could be written in $t$-ordered form as
	\begin{eqnarray}
	\label{sjt}
	{\left\{ {\hat F\left( {a^\dag,a} \right)} \right\}_{{s_2}}}{\left\{ {\hat G\left( {a^\dag,a} \right)}\right\}_{{s_3}}}\cdots 	&&{\left\{ {\hat H\left( {a^\dag,a} \right)} \right\}_{{s_k}}} = \nonumber \\ 								\sum\limits_{{\text{$i$-pair $\left({{u_l},t} \right)$-contractions}}}&&{{\left\{ {{\left( {\hat F\hat G \cdots \hat H} \right)}_i^{\mathbf{u}}\left( {a^\dag, a} \right)} \right\}}_t}~,
	\end{eqnarray}
in which $l \in \left\{ {1,2, \ldots ,i} \right\}$, ${\mathbf{u}} \equiv \left( {{{u}_1},{{u}_2} \ldots ,{{u}_i}} \right)$ with $u_l \in \left\{ {{-1,1,s_j}} \right\}$, and ${\left( {\hat F\hat G \cdots \hat H} \right)_i^{{\mathbf{u}}}}$ is the $i$-pair $\left (u_l,t\right )$-contracted form of the multiplicative sequence $\hat F\hat G \cdots \hat H$. A $\left (s_j,t\right )$-contraction is defined through the relative $s_j$-order of two boson operators $a$ and $a^\dag$ to be the procedure of removing that pair and producing a factor of $\frac{t - s_j}{2}$.~\cite{SB}

As a simple example, one may write the $s$-ordered form of the monomial ${\left( {{a^\dag }a} \right)^2}$ as
	\begin{eqnarray}
	{\left( {{a^\dag }a} \right)^2} &=& {\left\{ {{a^{\dag 2}}{a^2}} \right\}_s} + \left[ {\left( {\frac{{s + 1}}{2}} \right) +3\left( {\frac{{s - 1}}{2}} \right)} \right]{\left\{ {{a^\dag }a} \right\}_s} \nonumber \\
&+& \left( {\frac{{s + 1}}{2}} \right)\left({\frac{{s - 1}}{2}} \right) + {\left( {\frac{{s - 1}}{2}} \right)^2}~,
	\end{eqnarray}
which could be simply verified.

A special case of great interest to us is the one with $s_j = s$ and also $t=s$. In this case, all contraction inside the blocks vanish and we are left with the problem of counting the number contractions between mutual blocks. Based on this remark, in the next sections we illustrate the art of combining this new technique with combinatorial calculus to find the anti-normally ordered form of the operator $e^{\lambda a^\dag a}$ in almost two equivalent ways.

\section{Anti-normal Ordering of $\left( {a^\dag a} \right)^n$}

In the first step, consider the problem of anti-normal ordering the monomial ${\left( {{a^\dag }a} \right)^n}$. We do this using the special case of Eq.~\eqref{sjt} with $t=s_j = -1,~j \geqslant 2$ in two ways;

\begin{enumerate}
\item The first and the simplest analysis is as follows. Consider the monomial ${\left( {a{a^\dag }} \right)^{n+1}}$. Writing this as $n+1$ multiplicative blocks $\left( {a{a^\dag }} \right)\left( {a{a^\dag }} \right) \cdots \left( {a{a^\dag }} \right)$, one may simply realize that, using the consideration give in Sec.~\ref{CGCO}, the problem of anti-normally ordering these `anti-normally ordered' blocks is the same as the example given in Sec.~\ref{Intro} except for appearing a power of $-1$ due to $\left( 1,-1 \right)$-contractions. In this way, a $k$-pair $\left( 1,-1 \right)$-contraction is possible in $S \left( n+1,n+1-k \right)$ ways producing a factor of $\left( -1 \right)^k$ leading to
	\begin{equation*}
	{\left( {a{a^\dag }} \right)^{n + 1}} = \sum\limits_{k = 1}^{n + 1} {S\left( {n + 1,k} \right){{\left( { - 1} \right)}^{n + 1 - k}}{a^k}{a^{\dag k}}}~.
	\end{equation*}
Now, we may remove a factor $a$ from the left and a factor $a^\dag$ from the right of both sides. This simply gives
	\begin{equation} \label{Antin}
	{\left( {{a^\dag }a} \right)^n} = \sum\limits_{k = 1}^{n + 1} {S\left( {n + 1,k} \right){{\left( { - 1} \right)}^{n + 1 - k}}{a^{k - 1}}{a^{\dag k - 1}}}~,
	\end{equation}
which is the desired result.

\item Another way of solving the problem is to consider the monomial ${\left( {{a^\dag }a} \right)^n}$ as $n+1$ blocks of the form $\left( {{a^\dag }} \right)\left( {a{a^\dag }} \right) \cdots \left( {a{a^\dag }} \right)\left( a \right)$. All the blocks are again anti-normally ordered by themselves. Thus there exists no contraction inside the blocks and we just need to count the number of $k$-pair $\left( 1,-1 \right)$-contractions between $n+1$ mutual blocks which is given by $S \left( n+1,n+1-k \right)$. This leads to the result Eq.~\eqref{Antin}.

\end{enumerate}

Now, we may represent the anti-normally ordered form of the monomial ${\left( {{a^\dag }a} \right)^{n}}$ by means of Eq.~\eqref{Bell} as
	\begin{eqnarray} \label{AntiaDa}
	{\left( {{a^\dag }a} \right)^n} &=&  \vdots \frac{1}{{{a^\dag }a}}\sum\limits_{k = 1}^{n + 1} {S\left( {n + 1,k} \right)		{{\left({ - 1} \right)}^{n + 1 - k}}{a^k}{a^{\dag k}}}  \vdots  \nonumber \\ 
	&=& {\left( { - 1} \right)^{\left( {n + 1} \right)}} \vdots \frac{1}{{{a^\dag }a}}B\left( {n + 1, - {a^\dag }a} \right) \vdots~.
	\end{eqnarray}

\section{Operational Method to Anti-normal Ordering of $e^{\lambda a^\dag a}$}

As we have seen at the end of the previous section, we might formally use the actions like division by operators inside the ordering symbols to express the answers in terms of well-known polynomials or functions. In this section we extend this vision to combine the operational methods with our combinatorial approach to solve the problem of anti-normal ordering of the operator $e^{\lambda a^\dag a}$.

To this end we expand the exponential and use Eq.~\eqref{AntiaDa} to write
	\begin{eqnarray*}
	{e^{\lambda {a^\dag }a}} &=& \sum\limits_{n = 0}^\infty  {\frac{{{\lambda ^n}}}{{n!}}{{\left( {{a^\dag }a} \right)}^n}}  		\nonumber \\
	&=&  \vdots \frac{{ - 1}}{{{a^\dag }a}}\sum\limits_{n = 0}^\infty  {\frac{{{{\left( { - \lambda } \right)}^n}}}{{n!}}B\left( {n 	+ 1, - {a^\dag }a} \right) \vdots}~.
	\end{eqnarray*}
Now we may synchronize the summation index using the operation of $\frac{\partial }{{\partial \left( -\lambda \right)}}$ and write
	\begin{equation} \label{genaDa}
	{e^{\lambda {a^\dag }a}} =  \vdots \frac{1}{{{a^\dag }a}}\frac{\partial }{{\partial \lambda }}\sum\limits_{n = 0}^\infty  {\frac{{{{\left( { - \lambda } \right)}^{n + 1}}}}{{\left( {n + 1} \right)!}}B\left( {n + 1, - {a^\dag }a} \right)} \vdots~.
	\end{equation}
Substituting the definition of generating function of Bell polynomials Eq.~\eqref{Dobinski} into Eq.~\eqref{genaDa} gives
	\begin{eqnarray}
	{e^{\lambda {a^\dag }a}} &=& \vdots \frac{1}{{{a^\dag }a}}\frac{\partial }{{\partial \lambda }}\left[ {G\left( { - \lambda , - {a^\dag }a} \right) - 1} \right] \vdots  \nonumber \\ 
	&=&  \vdots \frac{1}{{{a^\dag }a}}\frac{\partial }{{\partial \lambda }}{e^{ {a^\dag }a\left( {1 - {e^{ - \lambda }}} \right)}} \vdots  \nonumber \\ 
	&=& {e^{ - \lambda }} \vdots {e^{{a^\dag }a\left( {1 - {e^{ - \lambda }}} \right)}} \vdots~,
	\end{eqnarray}
which is the same result as given in Ref.~\cite{MMQO}.

\section{Conclusion}

In summary, we have shown that the process of anti-normal ordering of some operator function might be gently solved using the combinatorial identities and simple operational methods. In fact, we have addressed a systematic combinatorial approach to the problem. We have exploited this technique to anti-normal the operator ${e^{\lambda {a^\dag }a}}$ as well as giving the procedure's combinatorial counterpart by two neat and clear analysis.

\end{document}